\documentclass[%
reprint,
superscriptaddress,
 amsmath,amssymb,amsfonts,
 aps,
prx
 citeautoscript,
floatfix,
longbibliography,
]{revtex4-2}

\usepackage[utf8]{inputenc}
\usepackage[english]{babel}
\usepackage[T1]{fontenc}
\usepackage{float}
\usepackage{graphicx}
\usepackage{dcolumn}
\usepackage{amsmath}
\usepackage{amssymb}
\usepackage{bm}
\usepackage{xcolor}
\usepackage{listings}
\usepackage{enumitem}
\usepackage{epstopdf}
\usepackage[normalem]{ulem}
\usepackage{hyperref}
\hypersetup{
    colorlinks=true,
    linkcolor=blue,
    citecolor=blue,
    filecolor=blue,      
    urlcolor=blue,
    }
\usepackage{physics}
\usepackage{braket}
\usepackage{comment}
\usepackage{subcaption}
\usepackage{caption}
\begin{document}

\newcommand{\cd}{\hat{c}^\dagger}
\newcommand{\cnd}{\hat{c}}
\newcommand{\sigmab}{\bar{\sigma}}
\newcommand{\bsh}{\bold{\hat{S}}}
\newcommand{\sh}{\hat{S}}
\newcommand{\up}{\uparrow}
\newcommand{\down}{\downarrow}


\title{Strengthened correlations near [110] edges of $d$-wave superconductors in the t-J model with the Gutzwiller approximation}

\author{A. Joki}
\email[e-mail: ]{amborn@chalmers.se}
\affiliation{Department of Microtechnology and Nanoscience - MC2,
Chalmers University of Technology,
SE-41296 G\"oteborg, Sweden}

\author{M. Fogelstr\"om}
\affiliation{Department of Microtechnology and Nanoscience - MC2,
Chalmers University of Technology,
SE-41296 G\"oteborg, Sweden}
\affiliation{
Nordita, KTH Royal Institute of Technology and Stockholm University, Hannes Alfv\'{e}ns v\"{a}g 12, 10691 Stockholm, Sweden
}

\author{T. L\"ofwander}
\affiliation{Department of Microtechnology and Nanoscience - MC2,
Chalmers University of Technology,
SE-41296 G\"oteborg, Sweden}

\date{\today}

\begin{abstract}

We report results of a study of strongly correlated $d$-wave superconducting slabs within a t-J model solved using the statistically consistent Gutzwiller approach. For different dopings, we model slabs cut at 45$^\circ$ relative to the main crystallographic $ab$-axes, i.e., [110] orientation, and conclude that quasiparticle charge is drawn to the edges.
Thus, the correlations are locally strengthened, and a region near the edge is closer to the Mott insulating state with reduced hopping amplitudes on the bonds. Superconductivity is locally weakened near the edges in this correlated state, and the spectral weight of zero-energy Andreev bound states is substantially reduced compared to weak coupling theory. The renormalization of the edge states leaves no room for an extended $s$-wave component to form for hole dopings ranging from strongly underdoped to slightly overdoped.

\end{abstract}

%
%

\maketitle

\section{Introduction}

Quantum materials and their use have attracted intense research in recent years \cite{TokKawNag17,BasAveHsi17,WanZha17,SatAnd17}. The material properties, emerging from strong correlations and topology, often guarantee protected states at the material surfaces. However, intricate physics near the surface may lead to instabilities and surprising readjustments of the surface states \cite{WanMeiGef17,AmaPriPet17}. One class of materials of broad interest is the cuprates, which are high-temperature superconductors with $d$-wave symmetry of the order parameter \cite{TsuKir00,Har95}. Surfaces not perfectly aligned with the main crystallographic axes of these $d$-wave superconductors host a large spectral weight of zero-energy Andreev bound states \cite{Hu94} related to topology \cite{RyuHat02,SatTanYad11}.

High-temperature superconductivity is believed to emerge from a Mott-insulating state upon hole doping and is characterized by strong electron-electron interactions \cite{LeeNagWen06}. The t-J model, together with the Gutzwiller approximation, has been used for a long time to study properties of strongly correlated superconductors \cite{ZhaGroRic88,EdeMutGro}.
This approximation scheme can reproduce many results of variational Monte Carlo methods which are in qualitative agreement with experiments \cite{AndLeeRan2004,RanSenTri12}.

Recent focus has shifted from bulk properties of cuprates, to the readjustment of the many-body state in response to inhomogeneities, such as defects \cite{TsuTanOga01,GarArtRan08,ChrHirAnd11,ChaGho14}, vortices \cite{TsuHirOga03,LiuTuChe23,MahBanPep25}, interfaces \cite{CheCheWan10}, grain boundaries \cite{WolGraLod12} and edges \cite{MatKon20,MatKon20b,ChaLofFog22}. These studies on inhomogeneous systems have mainly been within the framework of the t-J model and the Gutzwiller approximation. In one of these, superconductivity coexisting with antiferromagnetism, was observed near impurities together with a non-uniform redistribution of charge \cite{ChrHirAnd11}. A redistribution has also been seen in the studies in  Refs.~\cite{GarArtRan08,ChaGho14}. The redistribution effectively screens the superconductor from the adverse effect of impurity scattering and can contribute to explaining the robustness of the high-$T_c$ superconductors against disorder. Analogous redistribution of charge is seen in the other works on vortices and interfaces \cite{TsuHirOga03,LiuTuChe23,MahBanPep25,WolGraLod12,ChaLofFog22}.

 Redistribution of charge, has also been seen in inhomogeneous superconductors studied with the Bogoliubov-de Gennes approach including Hartree-Fock (HF) interactions. In Ref.~\cite{CheCroSha09}, the electron density across the radius of nanowires in the superconducting state was found to be non-uniform when HF terms were included. In another study of surface superconductor-insulator transitions under an applied electric field at the surface, it was seen that the HF interactions increase the effect of the applied field \cite{CheZhuZha24}.

Edges of $d$-wave superconductors misaligned with respect to the main crystallographic axes have attracted interest both experimentally and theoretically since the 1990s \cite{Sig98,KasTan00,LofShuWen01}. At these edges zero-energy Andreev bound states are formed due to the $d$-wave symmetry of the order parameter \cite{Hu94}. Experimentally, the Andreev bound states have been observed in transport measurements through normal metal-superconducting contacts as zero-bias conductance peaks (ZBCP) \cite{KasTan00}. In a magnetic field, the ZBCP split, an effect consistent with Doppler shifts due to screening supercurrents \cite{CovAprPar97,FogRaiSau97}. The same effect has been seen without applying a magnetic field \cite{CovAprPar97,DagDeu01}. This seemingly spontaneous split is in agreement with suggestions of time reversal symmetry breaking and appearance of a subdominant component of the order parameter with $s$-wave symmetry \cite{MatShi95,SigKubLee96,FogRaiSau97,TanTanOga99}. There are, however, other competing mechanisms that rely on magnetic ordering \cite{HonWakSig00,PotLee14,SejWalLof24} or spontaneous supercurrents \cite{Hig97,BarKalKur00,LofShuWen00,HakLofFog15}. Experimentally, signatures of spontaneous time-reversal symmetry breaking still appear to be scarce \cite{CovAprPar97,DagDeu01,CarPolKor00,NeiHar02,KirTsuAri06,SaaMorSal11,GusGolFog12}.

In a more recent study, spontaneous breaking of both translational and time-reversal symmetries was proposed \cite{HakLofFog15,HolVorFog18,Holmvall:2020,WalAskHol20}. The translational symmetry breaking is manifested through supercurrent loops near the edges. This phase was shown to also appear in a strongly correlated t-J model within the Gutzwiller inhomogeneous mean field theory \cite{ChaLofFog22}. The study was done for a hole doping of $\delta=0.2$ and showed some enhanced electron occupation near the edge and an enhanced robustness of the symmetry broken phase to disorder compared to weak coupling theory. In the present paper, we focus on the doping dependence of the redistribution of charge and its effect on slabs with pair-breaking [110] edges. Assuming translational invariance, we neglect the possible formation of spontaneous current loops, and study how strengthened correlations affect the superconducting order parameter and the Andreev bound states themselves, as well as the fate of a subdominant extended $s$-wave order parameter.

The paper is organized as follows. In Section~\ref{sec:tJmodel} the t-J model and the statistically consistent Gutzwiller approach (SGA) is presented, including the main steps needed to derive the equations applicable to a slab with edges misaligned 45$^\circ$ with respect to the main crystallographic axes. The results are presented in Section~\ref{sec:results}. In particular, Section~\ref{sec:ResultsHomogeneous} contains a comparison of the results obtained here to previous results \cite{TanTanOga99} where electron densities, and resulting renormalizations due to strong correlations, were assumed to retain their bulk values throughout the slab. The present paper is summarized in Section~\ref{sec:Summary} and some technical details have been collected in the Appendix.

\section{\protect\lowercase{t}-J model and statistically consistent Gutzwiller approach}\label{sec:tJmodel}

The starting point for the investigation is the t-J Hamiltonian \cite{ChaSpaOle77, ZhaGroRic88,EdeMutGro}
\begin{equation}
\hat{H}
= t \sum_{\braket{ij}\sigma} \hat{P}(\cd_{i\sigma}\cnd_{j\sigma}
+ \mathrm{h.c.})\hat{P}
+ J \sum_{\braket{ij}}\bsh_i\cdot{ \bsh_j},
\label{eq:tJ-projected}
\end{equation}
where $i$ and $j$ label sites on a square lattice in two dimensions, $\sum_{\braket{ij}}$ means sum over nearest neighbors, $t<0$ is the nearest neighbor hopping amplitude, and $J>0$ is the antiferromagnetic superexchange coupling. The index $\sigma \in \{\up, \down\}$ labels spin. The operator $\cd_{i\sigma}$ creates and $\cnd_{i\sigma}$ annihilates an electron with spin $\sigma$ on site $i$, while $\bsh_i$ is the spin operator. The Gutzwiller projection operators $\hat{P}$ forbid changes of the number of doubly occupied sites \cite{Gut63}. The projection operators are dealt with using the Gutzwiller approximation \cite{Gut63,OgaTohKan75,Voll84,EdeMutGro,ZhaGroRic88,OgaHim02,KacSpa11,AbrKacSpa13}, relating expectation values of an operator $\hat{O}$ evaluated in the correlated state $\braket{\Psi| \hat{O}| \Psi}$ to the corresponding expectation value evaluated in the uncorrelated state $\braket{\Psi_0| \hat{O} |\Psi_0}$ via a Gutzwiller factor $g^O$, unique for each operator
\begin{equation}
\frac{\braket{\Psi| \hat{O}| \Psi}}{\braket{\Psi|\Psi}}\approx g^O \frac{\braket{\Psi_0| \hat{O} |\Psi_0}}{\braket{\Psi_0|\Psi_0}}.
\end{equation}
In a search for the minimum ground state energy, the expectation value of the t-J Hamiltonian, Eq.~\eqref{eq:tJ-projected}, is computed utilizing the Gutzwiller approach. An effective Hamiltonian without projection operators is then introduced
\begin{equation}
\hat{H}_\mathrm{eff}
= t \sum_{\braket{ij}\sigma} (g_{ij}^t\cd_{i\sigma}\cnd_{j\sigma}
+ \text{h.c.})
+ J \sum_{\braket{ij}} g_{ij}^s\bsh_i\cdot{ \bsh_j},
\label{eq:tJ-Gutzwiller}
\end{equation}
with Gutzwiller factors renormalizing the hopping amplitude, $t\rightarrow tg_{ij}^t$, as well as the antiferromagnetic superexchange coupling, $J\rightarrow Jg_{ij}^s$. These factors depend on the local occupations of nearest neighbor sites as \cite{ZhaGroRic88}
\begin{align}
\begin{split}
g^t_{ij} &= 2\sqrt{\frac{1-n_i}{2-n_i}\frac{1-n_j}{2-n_j}},\\
g^s_{ij} &= \frac{4}{(2-n_i)(2-n_j)}.
\label{eq:Gutzwiller_factors}
\end{split}
\end{align}

The expectation value $\braket{\Psi_0| \hat{H}_\mathrm{eff}|\Psi_0}$ is evaluated using a mean-field approximation. We have considered mean fields for hopping and superconductivity, but also allowed for antiferromagnetic order by computing inhomogeneous and spin-dependent occupations $n_{i\sigma}$. As soon as antiferromagnetism is present, superconductivity becomes a mix of a spin-singlet component and a spin-triplet with zero spin projection. However, in the hole doping region $\delta\in[0.05,0.2]$ we are considering here, we have found that antiferromagnetism is not stabilized and it will not be considered from here on. Thus, we define mean fields in the uncorrelated (unprojected) state for hopping and pairing as
\begin{align}
\begin{split}
\chi_{ij}   &= \braket{\cd_{i\sigma}\cnd_{j\sigma}}_0 
             = \chi_{ji}^*,\\
\Delta_{ij} &=  \braket{\cnd_{i\down}\cnd_{j\up}}_0
             = -\braket{\cnd_{i\up}\cnd_{j\down}}_0,
\label{eq:mean_fields_definition}
\end{split}
\end{align}
where the last equality on the first line is a result of the fermionic commutation relations, while on the second line it derives from the spin-singlet symmetry of the considered superconductivity. The expectation value of the effective Hamiltonian in Eq.~\eqref{eq:tJ-Gutzwiller} then takes the form 
\begin{align}
\begin{split}
\braket{\hat{H}_\mathrm{eff}}_0\approx W=\sum_{\braket{ij}}\left[4tg_{ij}^t \chi_{ij}-\tfrac{3}{2}Jg_{ij}^s\left(|\chi_{ij}|^2+|\Delta_{ij}|^2\right)\right].
\label{eq:effective_energy}
\end{split}
\end{align}
Following the statistically consistent Gutzwiller approach (SGA) \cite{JedSpa10,KacSpa11,AbrKacSpa13}, the mean-field Hamiltonian is defined
\begin{align}
\begin{split}
\hat{K} &= W - \sum_{i} \left[ \lambda_{i}^n(\hat{n}_{i}-n_{i}) + \mu\hat{n}_{i} \right]\\
&- \sum_{\braket{ij}\sigma} \left[
       \lambda_{ij}^\chi(\cd_{i\sigma}\cnd_{j\sigma}-\chi_{ij})
     + \lambda_{ij}^\Delta(\cnd_{i\sigmab}\cnd_{j\sigma}-\Delta_{ij})
     + \mathrm{h.c.}\right],
\end{split}
\label{eq:GrandHamiltonian}
\end{align}
where $\mu$ is the chemical potential, and Lagrange multipliers $\lambda$, one for each mean-field as well as for local occupation numbers, have been introduced. The Lagrange multipliers follow the same symmetry constraints as the corresponding mean-fields.

The grand potential functional at inverse temperature $\beta=1/k_\mathrm{B} T > 0$ is then defined as
\begin{equation}
\Omega = -\frac{1}{\beta}\ln{\text{Tr}\left(e^{-\beta \hat{K}}\right)},
\end{equation}
where $k_\mathrm{B}=1$ is the Boltzman constant. The goal of the calculation is to find the minimum $ \Omega_0$ of the grand potential functional with respect to the mean fields and the constraints introduced through the Lagrange multipliers. The procedure will be outlined below after introducing the specific geometry of the slab. 

\subsection{Slab with [110] edges}\label{sec:slab}

\begin{figure}[b]
\includegraphics[width=1\linewidth]{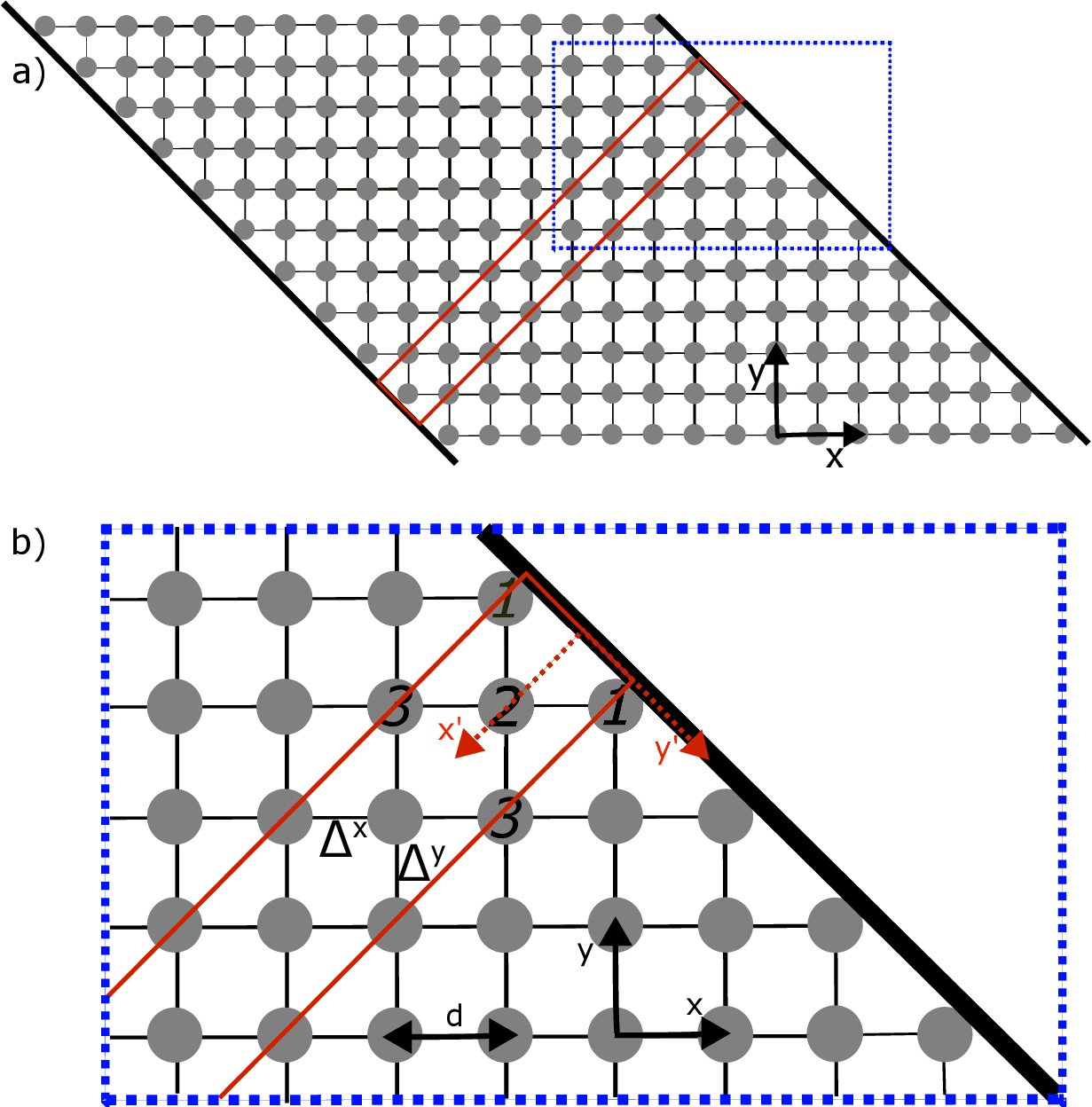}
\caption{(a) Square lattice with lattice constant $d$ cut into a slab with [110] edges. There is translational invariance along the edge tangent. The red lines mark the unit cell, the gray disks sites, and the $x$- and $y$-axes are the crystal axes. Lines between sites represent the links in either $x$- or $y$-direction.
(b) The area enclosed by dotted blue lines in (a), including the right [110] edge with edge tangent $y'$ and edge normal $x'$. The sites are labelled $a=1$, $2$, $3$, etc. dependent on distance from the edge measured along $x'$. Note that between sites $a$ and $a+1$ there will be one link in the $y$-direction and one in the $x$-direction. The pairing amplitudes $\Delta^x_{a,a+1}$ and $\Delta^y_{a,a+1}$, laying on links, are computed independently in order to allow for $d$-wave and extended $s$-wave symmetries.}
\label{fig:Lattice110}
\end{figure}

We consider a crystal with a 2-dimensional square lattice structure with lattice constant $d$ and focus on a slab with the two edges cut at 45$^\circ$ relative to the crystallographic axes ([110] edges), as illustrated in Fig.~\ref{fig:Lattice110}. The slab is translationally symmetric along the $y'$-axis (edge tangent), but is allowed to vary along the $x'$-axis measuring distance from the edge. The unit vectors in the rotated coordinate system with origo on the upper slab edge as in Fig.~\ref{fig:Lattice110}(b) are given by $\hat{x}'=-(\hat{x}+\hat{y})/\sqrt{2}$ and $\hat{y}'=(\hat{x}-\hat{y})/\sqrt{2}$. The slab has $N_x$ sites from edge to edge, and we shall use the integer $a=1,\,2,\dots,N_x$ to enumerate the sites as indicated for the first few sites in Fig.~\ref{fig:Lattice110}b. The red rectangle shows the unit cell, where sites with odd numbers lie with half weight on the unit cell edges while sites with even numbers lie along the unit cell center axis.
To compute the mean field  Hamiltonian $\hat K_\mathrm{110}$ for the slab geometry, we make use of the translational symmetry in the $\hat{y}'$-direction by considering $N_y$ unit cells and utilize a periodic boundary condition. The full calculation is presented in the Appendix and here we outline the main steps.

The expectation value of the effective Hamiltonian in Eq.~\eqref{eq:tJ-Gutzwiller} is, after summing over nearest neightbors, reduced to a sum over transverse sites
\begin{align}
\begin{split}
W_{110} &= N_y \sum_{a=1}^{N_x-1} \left[8t g_{a,a+1}^t \chi_{a,a+1}\right.\\
&\left. -\tfrac{3}{2} J g_{a,a+1}^s\left( 2|\chi_{a,a+1}|^2 + |\Delta_{a,a+1}^x|^2 + |\Delta_{a,a+1}^y|^2 \right)\right].
\end{split}
\label{eq:W110}
\end{align}
We note that, for instance, between sites $a=2$ and $a=1$ in Fig.~\ref{fig:Lattice110} there are two bonds, one directed in the $x$-direction and the other in the $y$-direction. We keep pairing amplitudes along the $x$- and $y$-directions in the unit cell separate to allow for both extended $s$-wave and $d$-wave symmetries. On the other hand, the hopping mean fields are assumed equal along these two bonds.

The terms with Lagrange multipliers and the creation and annihilation operators in Eq.~\eqref{eq:GrandHamiltonian} are Fourier transformed. An annihilation operator, $\cnd_{ak\sigma}$, is then labeled by an index $a$ indicating the transverse lattice site, a $k$-vector specifying the plane wave in the $y'$-direction, and spin $\sigma$. For each transverse site we introduce electron-hole space
\begin{equation}
\hat\psi_{ak} = \begin{pmatrix}
    \cnd_{ak\uparrow}\\
    \cd_{a-k\downarrow}
\end{pmatrix}.
\label{eq:vec_elc_hole}
\end{equation}
For the whole slab we define a vector $\hat{\Psi}^\dagger_k=(\hat{\psi}^\dagger_{1k},\hat{\psi}^\dagger_{2k},\dots,\hat{\psi}^\dagger_{N_x k})$ and write the full grand potential in terms of a $2N_x\times 2N_x$ large Hamiltonian matrix $\bm{H}_k$ as
\begin{equation}
\hat{K}_{110} = W_{110} + \lambda_{110} + \sum_k \hat{\Psi}_k^\dagger \bm{H}_k \hat{\Psi}_k,
\label{eq:GranPotantial110}
\end{equation}
where $\lambda_{110}$ consists of scalar terms, for example $\lambda_i^n n_i$ in Eq.~\eqref{eq:GrandHamiltonian}, as well as terms appearing after using fermionic commutation relations, see the Appendix for details. The Hamiltonian matrix $\bm{H}_k$ is band diagonal, where neighboring sites $a$ and $a+1$ are coupled through a $4\times 4$ block
\begin{widetext}
\begin{align}
\begin{split}
\left(\bm{H}_{k}\right)_{a,a+1} =
    \begin{pmatrix}
      -\lambda^n_{a}-\mu & 0 & -2\lambda^\chi_{a,a+1}\cos{\frac{kd}{\sqrt{2}}} & \eta^{-}_{a,a+1}(k)^*\\
          0 & \lambda^n_{a}+\mu & -\eta_{a,a+1}^+(k) & 2\lambda^\chi_{a,a+1}\cos{\frac{kd}{\sqrt{2}}} \\
          -2\lambda^\chi_{a,a+1}\cos{\frac{kd}{\sqrt{2}}} & -\eta_{a,a+1}^+(k)^* & -\lambda^n_{a+1}-\mu  & 0\\
          \eta_{a,a+1}^-(k)  & 2\lambda^\chi_{a,a+1}\cos{\frac{kd}{\sqrt{2}}}&  0 & \lambda^n_{a+1}+\mu 
    \end{pmatrix}.
\end{split}\label{eq:H_k}
\end{align}
\end{widetext}
where $\eta_{aa'}^{\pm}(k) = \lambda^{\Delta^x}_{aa'}e^{\pm ikd/\sqrt{2}} + \lambda^{\Delta^y}_{aa'}e^{\mp ikd/\sqrt{2}}$. The Hamiltonian $\bm{H}_k$ can be diagonalized, corresponding to the Bogoliubov transformation in a weak coupling Bogoliubov-de Gennes approach. We then obtain eigenvalues and eigenvectors specifying the new basis, in which the grand potential functional can be evaluated to be
\begin{equation}
\Omega = W_{110} + \lambda_{110} - \frac{1}{\beta}\frac{1}{N_y}\sum_{k}\sum_{l=1}^{2N_x}\ln{(1+e^{-\beta E_{lk}})},
\end{equation}
where we have $2N_x$ eigenvalues $E_{lk}$ for each wave vector $k$, enumerated by the index $l$. The eigenvalue spectrum is symmetric with respect to the chemical potential $\mu$ and forms a bandstructure $E_{lk}$ with respect to the wave vector $k$ in the first Brillouin zone $k\in\,\left]-\pi d/\sqrt{2}, \pi d/\sqrt{2}\,\right]$. Often we shall consider the grand potential functional per site $\Omega/N$, where $N=N_x N_y$.

The above set of equations is the starting point for the minimization of the grand potential functional, which is done as follows. First, searching for the minimum of $\Omega$ with respect to the mean-fields, we obtain equations for the Lagrange multipliers, see Eqs.~\eqref{eq:lambdas} and~ \eqref{eq:lambda_n}. Second, searching for the minimum of $\Omega$ with respect to the Lagrange multipliers, we obtain self-consistency equations for the mean fields, see Eq.~\eqref{eq:mean_fields}, which involves computing the derivatives of the eigenvalues $E_{lk}$ with respect to the Lagrange multipliers. 

The equations are numerically solved in an iterative manner until self-consistency has been achieved. We use $|t|=1$ as the unit of energy. The parameters are taken as $t=-1$, $J=0.25$, and inverse temperature $\beta=500$. The target average hole doping of the slab
\begin{equation}
\delta = 1 - \frac{1}{N_x}\sum_a n_a,
\end{equation}
is enforced by self-consistently tuning the slab chemical potential $\mu$.

\subsection{Local density of states}
The Gutzwiller approximation also affects the local density of states (LDOS) \cite{GarArtRan08,ChrHirAnd11,Fuk08}. The LDOS at site $a$ as function of energy $\omega$ is given by
\begin{equation}
\mathcal{N}_{a}(\omega) = \sum_{lk} g_{a}^\mathcal{N}\left[|u_{alk}|^2\delta(\omega-E_{lk})+|v_{alk}|^2\delta(\omega+E_{lk})\right],
\label{eq:ldos}
\end{equation}
where $l$ counts the positive eigenvalues $E_{lk}$. The spectral weights are given by the eigenvectors $u_{lak}$ (related to the electron-like part), and $v_{lak}$ (related to the hole-like part) of $\mathbf{H}_k$. The Gutzwiller factor is 
\begin{equation}
g_{a}^\mathcal{N}=g_{aa}^t=\frac{2(1-n_a)}{2-n_a}.
\end{equation}
The delta functions are approximated by Lorentzian functions with width $\Gamma$.

\section{Results}\label{sec:results}
\begin{figure}[b]
    \centering
    \includegraphics[width=\linewidth]{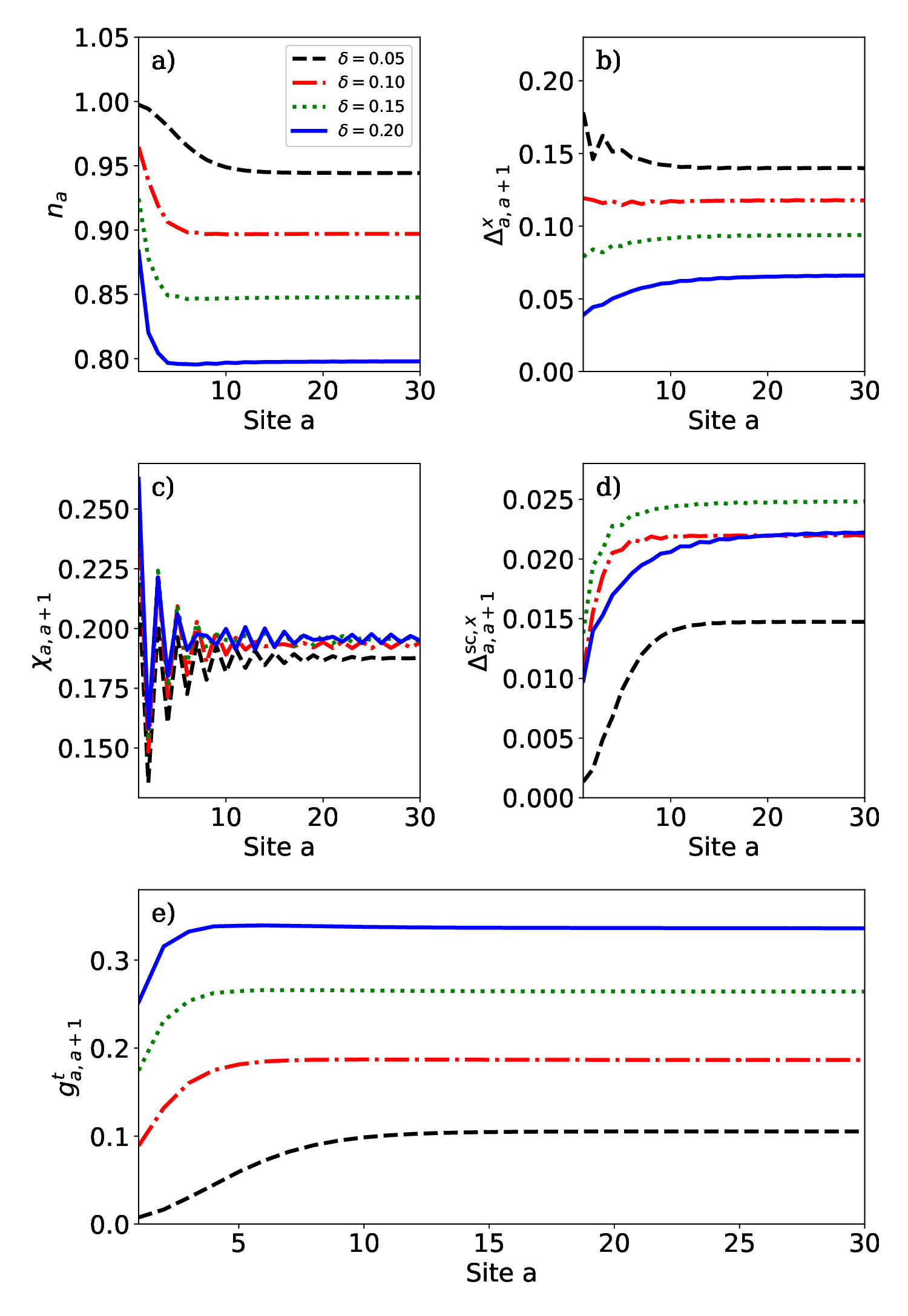}
    \caption{ (a) Spatially dependent electron densities $n_a$, (b) pairing amplitude $\Delta^x_{a,a+1}$, (c) hopping amplitude $\chi_{a,a+1}$, (d) superconducting order parameter $\Delta^{\mathrm{sc},x}_{a,a+1}$, and (e) the Gutzwiller factor $g^t_{a,a+1}$. All quantities are plotted at four different average hole dopings given in the legend. The slab width is $N_x=100$, but only the first $a\in[1,30]$ sites/links are shown. }
\label{fig:mean_fields}
\end{figure}
We present results for the spatial dependences of the electron density $n_a$, the pairing $\Delta_{a,a+1}^x$ and hopping $\chi_{a,a+1}$ mean fields in the uncorrelated state, the superconducting mean field in the correlated state, or superconducting order parameter,
\begin{equation}
\Delta^{\mathrm{sc},x}_{a,a+1}=g^t_{a,a+1}\Delta_{a,a+1}^x,
\label{eq:order_parameter}
\end{equation}
and the Gutzwiller factor $g^t_{a,a+1}$
in Fig.~\ref{fig:mean_fields}(a)-(e),  respectively, for a $N_x=100$ sites wide slab. The results are obtained for four average hole dopings in the interval $\delta\in[0.05, 0.2]$. The result of the calculation is a pairing amplitude of pure $d$-wave symmetry, $\Delta_{a,a+1}^y=-\Delta_{a,a+1}^x$. For simplicity we therefore only plot mean fields on $x$-links. 

We see in Fig.~\ref{fig:mean_fields}(b) that for high average hole doping, $\delta=0.2$, the pairing amplitude $\Delta^x_{a,a+1}$ is suppressed on the scale of the superconducting coherence length close to the edge, as expected for $d$-wave superconductivity near a [110] edge. At lower average hole dopings, the suppression of $\Delta^x_{a,a+1}$ is not as significant. At $\delta=0.05$, it is actually strengthened, reflecting enhanced electron-electron interactions pushing the edge towards the Mott insulating state. However, as shown in Fig.~\ref{fig:mean_fields}(d), the superconducting order parameter $\Delta^{\mathrm{sc},x}_{a,a+1}$ is reduced at the edge for all hole dopings, especially also for $\delta=0.05$. This difference between the mean field in the uncorrelated state, $\Delta_{a,a+1}^x$, and superconducting order in the correlated state, $\Delta_{a,a+1}^{\mathrm{sc},x}$, is due to the increased electron density at the edge, see Fig.~\ref{fig:mean_fields}(a), reducing the value of the Gutzwiller factor, $g^t_{a,a+1}$, see Fig.~\ref{fig:mean_fields}(e). At the lowest hole doping $\delta=0.05$, the edge electron density reaches values close to unity, black dashed line in Fig.~\ref{fig:mean_fields}(a), and the edge is effectively close to Mott insulating. For the hopping mean field $\chi_{a,a+1}$, there are bond-to-bond oscillations near the edge, but the amplitude of the oscillations have no qualitative dependence on average hole doping $\delta$. Note that $\chi_{a,a+1}^x-\chi_{a,a+1}^y=0$ by assumption.

\begin{figure}
    \centering
    \includegraphics[width=\linewidth]{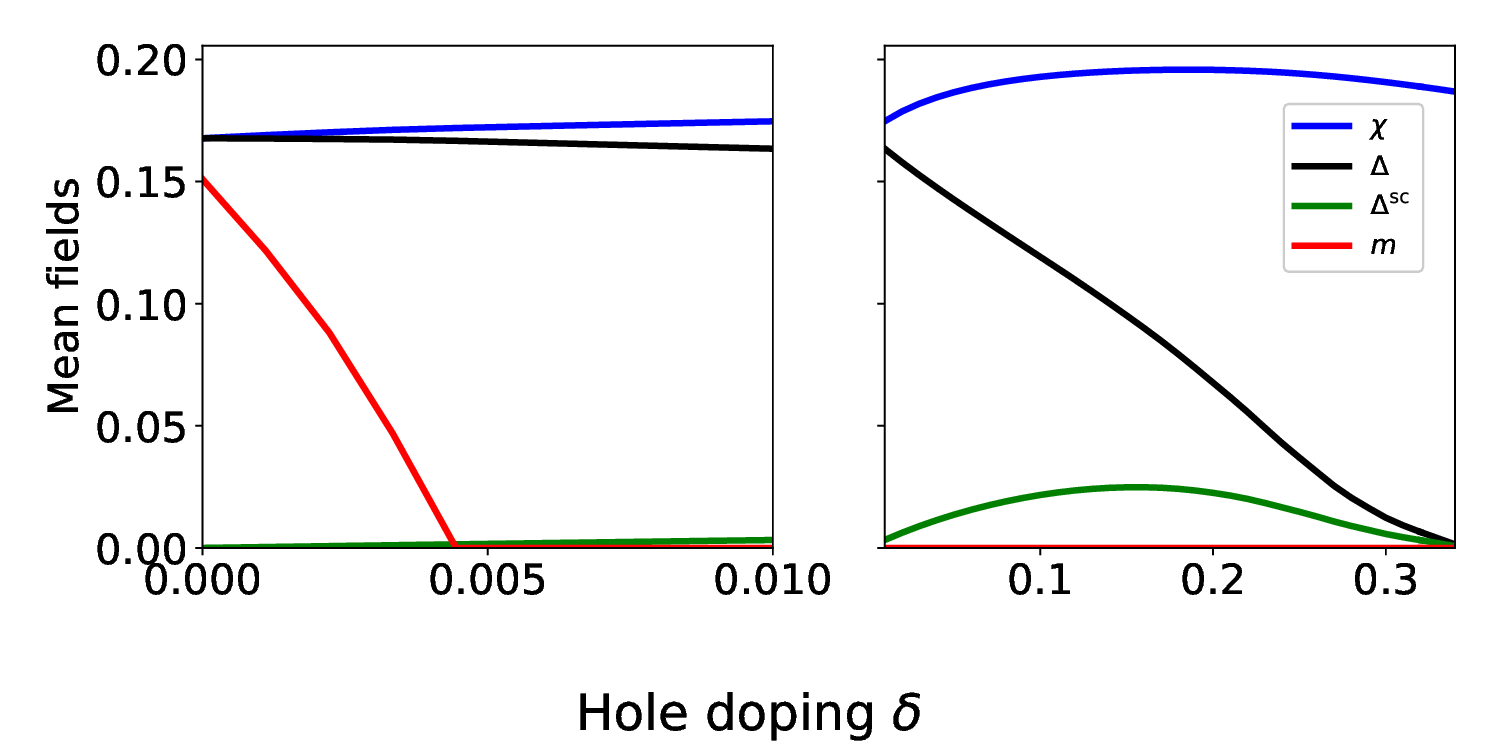}
    \caption{Doping dependence of self-consistently computed mean fields and superconducting order in a bulk system. The lowest hole dopings are shown in the left panel, where also antiferromagnetic order $m=n_\uparrow-n_\downarrow$ coexists with superconductivity in this model.}
    \label{fig:MomentumSpace}
\end{figure}

In Fig.~\ref{fig:MomentumSpace} we show for comparison the doping dependence of the pairing amplitude, hopping mean field, and superconducting order parameter for a bulk system computed with the same t-J model as above, but in reciprocal space, as in e.g. Ref.~\cite{AbrKacSpa13}. As is well known \cite{ZhaGroRic88}, the pairing amplitude $\Delta$ is increasing towards low doping, while the superconducting order parameter $\Delta^\mathrm{sc}$ shows a dome shape. Our results for the slab system above, is reflecting this behavior, where the edge regions of high electron density corresponds to the region with low $\delta$ in Fig.~\ref{fig:MomentumSpace}.

\begin{figure}
    \centering
    \includegraphics[width=\linewidth]{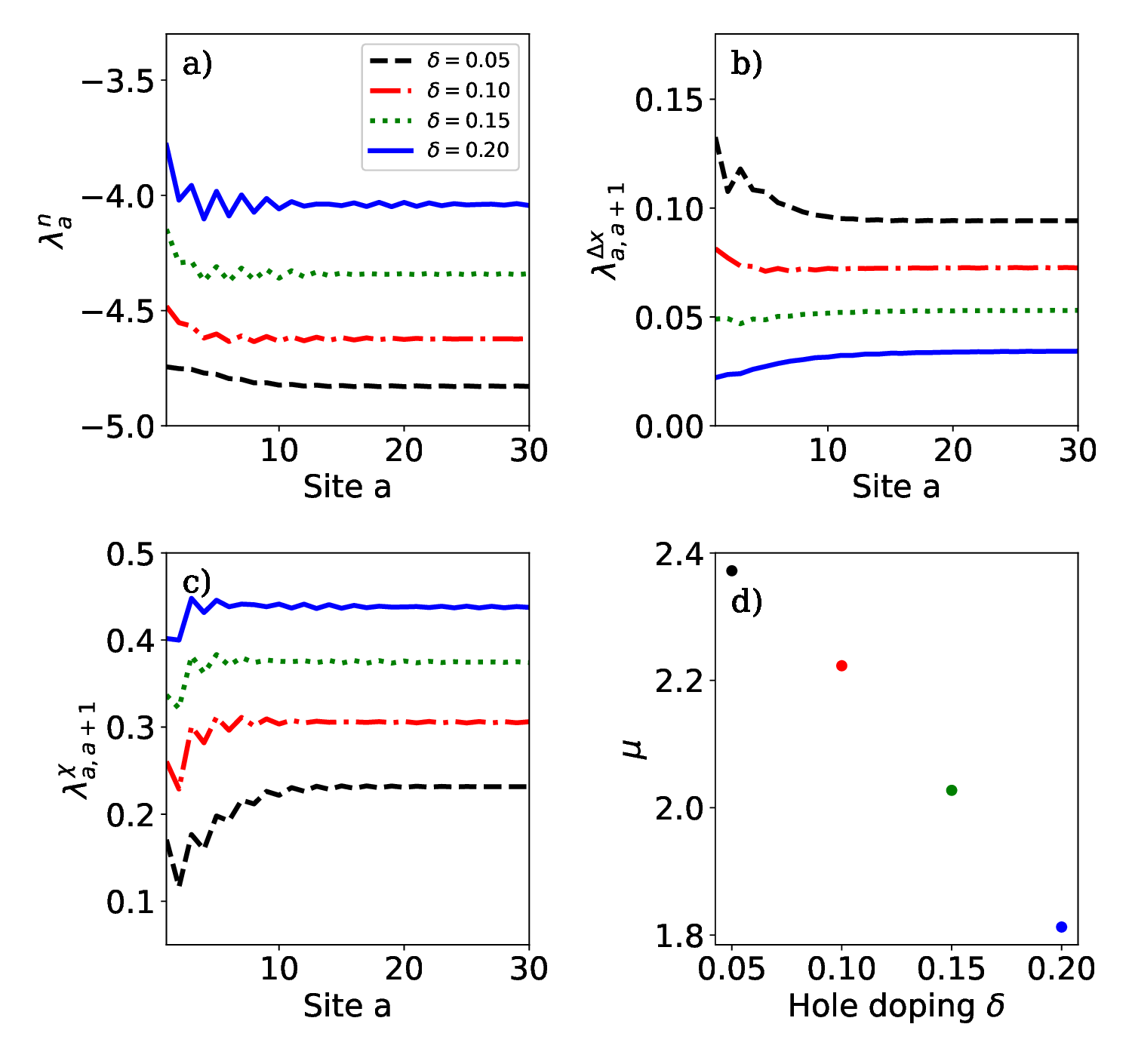}
    \caption{Spatial dependences of the Lagrange multiplyers for (a) local electron density, (b) $d$-wave pairing correlations, and (c) nearest neighbor hopping. Results for four average hole doping levels, indicated in the legend, are given. In (d) we show the doping dependence of the slab chemical potential with dopings on the x-axis.}
    \label{fig:lambda}
\end{figure}

In Fig.~\ref{fig:lambda} we show the spatial dependences of the Lagrange multipliers. Once self-consistency has been achieved they follow from Eqs.~(\ref{eq:lambdas})-(\ref{eq:lambda_n}). They consist of a renormalization of the hopping parameter $t$ and superexchange $J$ through local density dependent Gutzwiller factors, as well as Hartree and Fock shifts from the mean fields.
The Lagrange multipliers play an important role as they enter the Hamiltonian matrix $\mathbf{H}_k$ that is diagonalized to give eigenvalues and eigenvectors that determine the local densities, the mean fields, as well as the local density of states.

The Lagrange multiplier for local electron density $\lambda_a^n$ enters together with the slab chemical potential in Eq.~(\ref{eq:H_k}) and reflects the local enhancement near the edge of the electron density from electron-electron interactions, as seen in Fig.~\ref{fig:mean_fields}(a). It is weakly strengthened near the edge. On the other hand, $\lambda_{a,a+1}^\chi$ and $\lambda_{a,a+1}^{\Delta^x}$ enter in Eq.~\eqref{eq:H_k} in the positions for nearest neighbor hopping and superconducting correlations. Near the edge, the pairing correlations $\lambda_{a,a+1}^{\Delta^x}$ are strenthened for low doping $\delta$, while the renormalized hopping $\lambda_{a,a+1}^\chi$ is reduced for all $\delta$ considered here. In conclusion, the edge become more strongly correlated than the slab interior while the condensate, actual superconducting order defined in Eq.~\eqref{eq:order_parameter}, is locally weakened.

\subsection{Local density of states}\label{sec:ResultsLDOS}

\begin{figure} [t!]
    \centering
    \includegraphics[width=\linewidth]{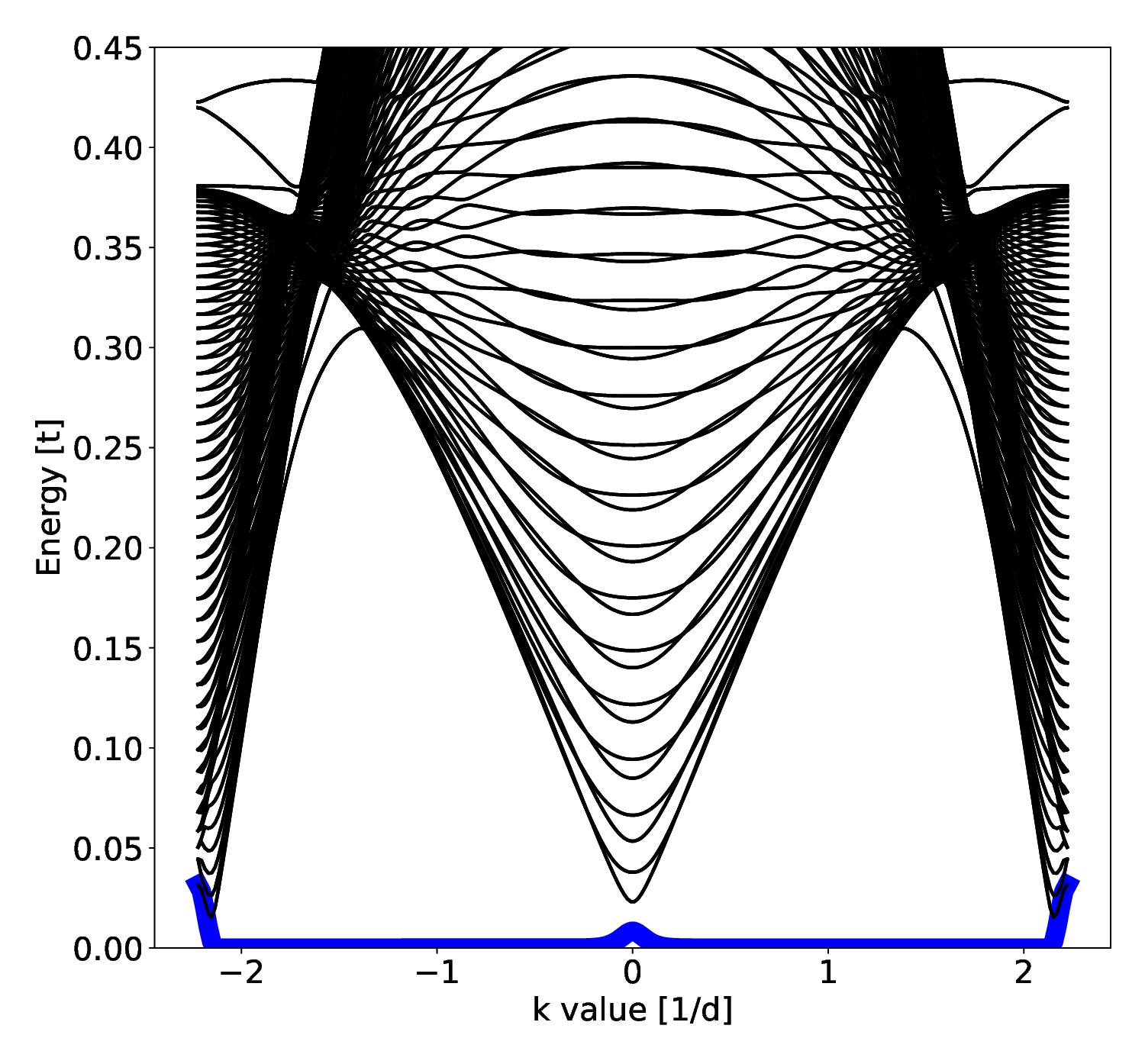}
    \caption{The spectrum of eigenvalues $E_{lk}$ for a $N_x=100$ wide slab at average hole doping $\delta=0.05$. The spectrum contains 100 bands at positive energies within the first Brillouin zone $k\in[-\pi d/\sqrt{2},\pi d/\sqrt{2}]$. Note that the spectrum is cut at an energy $0.450
    $. The thick blue line is the lowest positive eigenvalue and contains the zero-energy Andreev bound states.}
    \label{fig:eigenvalues}
\end{figure}

Next we study the eigenvalue spectrum and local density of states, focusing on low average hole doping $\delta=0.05$. In Fig.~\ref{fig:eigenvalues} the eigenvalue spectrum $E_{lk}$ is shown. The main feature is a flat band of zero-energy states highlighted as a bold blue line. These states follow from the $d$-wave symmetry of the superconductor and are associated with suppression of $d$-wave superconductivity near the edge, as studied in a large volume of literature within quasiclassical theory and Bogoliubov-de Gennes weak coupling theory \cite{Sig98,KasTan00,LofShuWen01}. Here we see that the actual zero-energy flat band survives when strong correlations are taken into account, although neither the $d$-wave pairing amplitude $\Delta_{a,a+1}^x$, nor the $d$-wave correlations $\lambda_{a,a+1}^{\Delta^x}$, are suppressed as reported above. Instead, the charging of the edge dominates these quantities.

\begin{figure}[t!]
    \centering
    \includegraphics[width=1\linewidth]{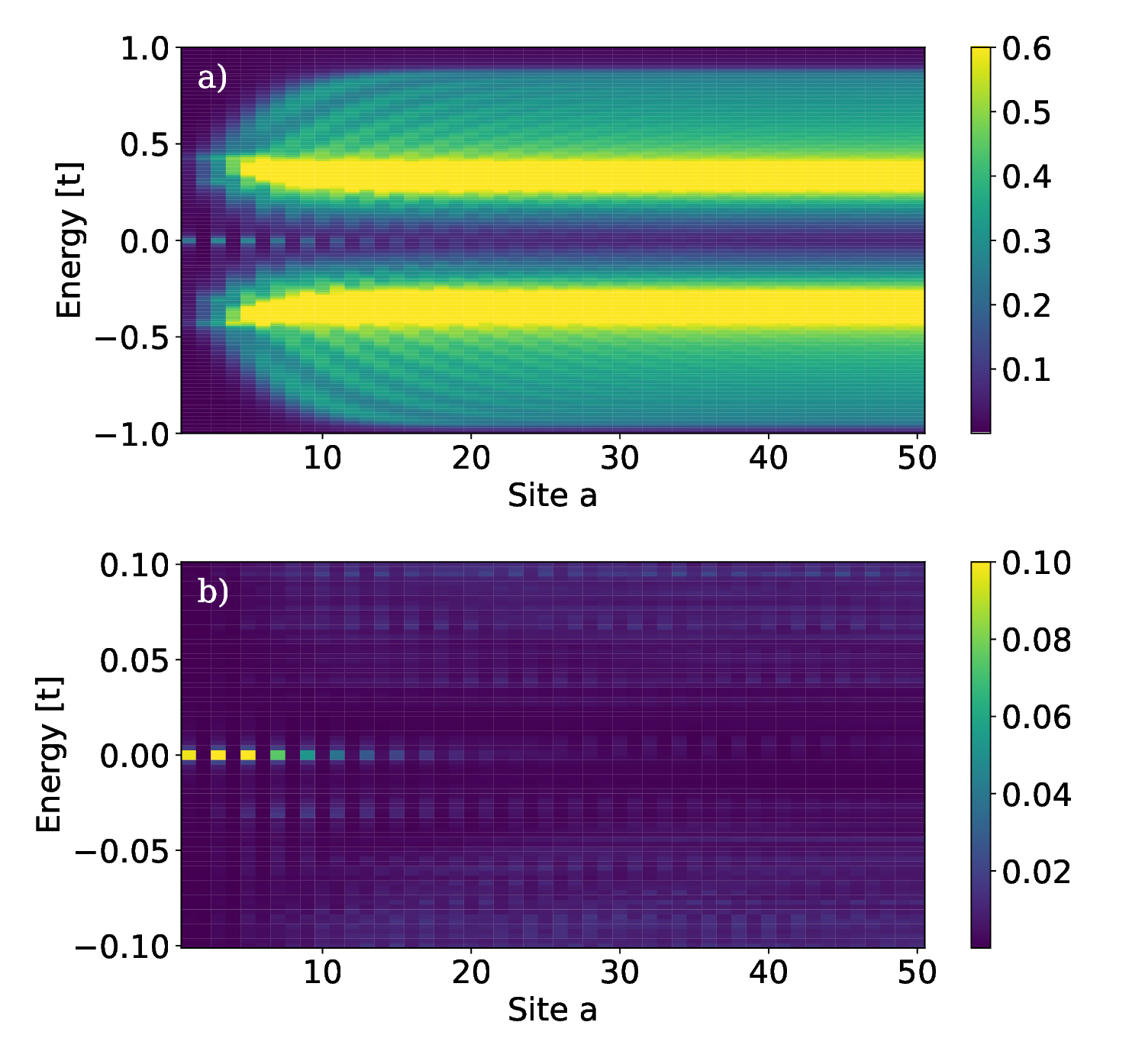}
     \caption{Local density of states $\Gamma\mathcal{N}_a(E)$, scaled by the Lorentzian width $\Gamma$ of the delta functions in Eq.~\eqref{eq:ldos}, for the same slab as in Fig.~\ref{fig:eigenvalues} ($N_x=100$, $\delta=0.05$). In (a) we show the full spectrum as function of coordinate $a$ using $\Gamma=0.015$. (b) Focus on the low-energy part of the spectrum, where the zero-energy Andreev bound states are clearly seen with a smaller $\Gamma=0.001$. 
    \label{fig:ldos}}
\end{figure}

\begin{figure}
    \centering
    \includegraphics[width=1\linewidth]{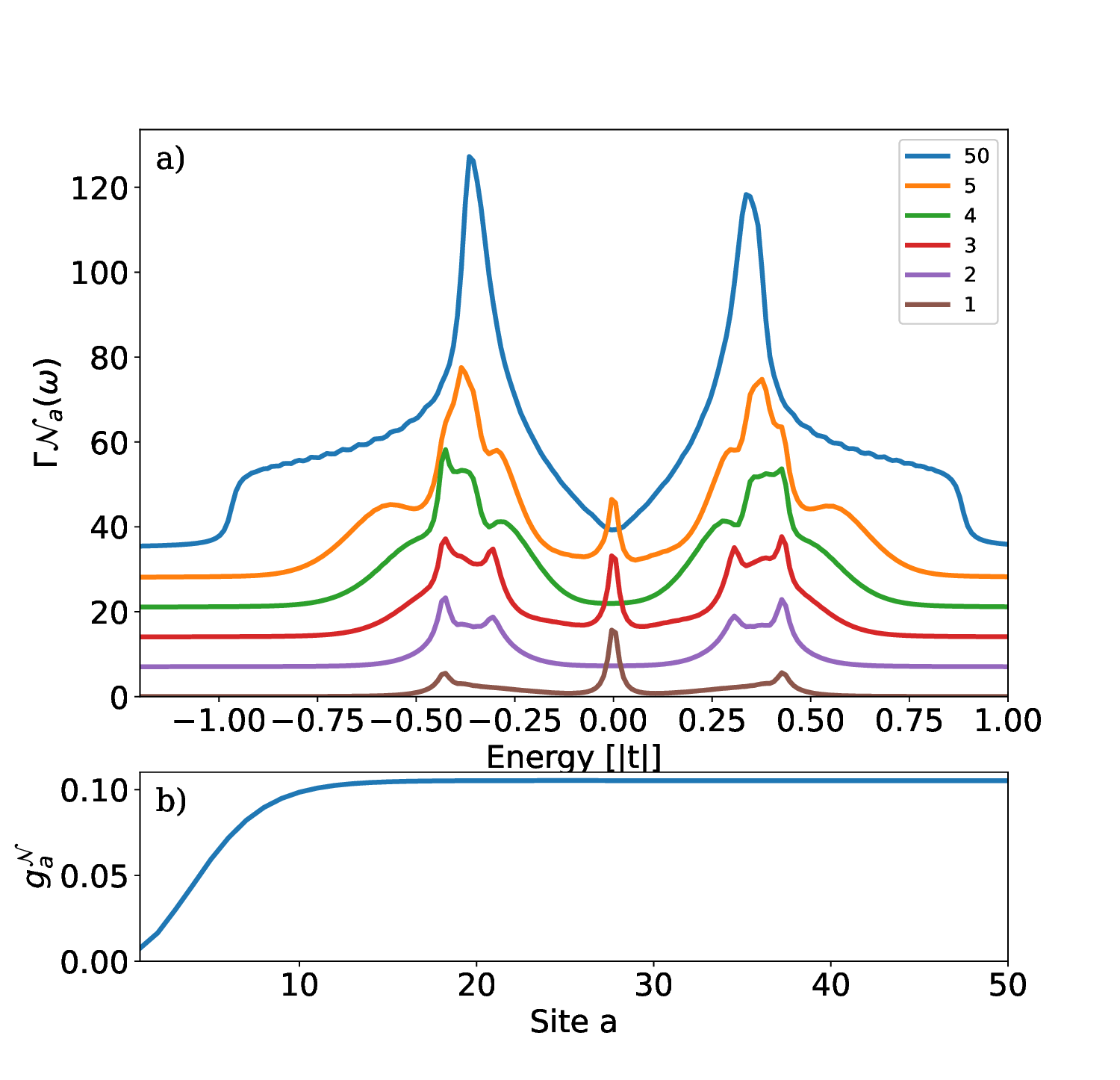}
     \caption{(a) Cuts of the local density of states $\Gamma\mathcal{N}_a(E)$ in Fig.~\ref{fig:ldos} ($N_x=100$, $\delta=0.05$) at sites between 1 and 50. Each consecutive curve is shifted upwards by 7 for clarity. (b) The spatial dependence of the Gutzwiller factor $g^N_a$ that enters into the LDOS defined in Eq.~\eqref{eq:ldos}.}
    \label{fig:ldos_cuts}
\end{figure}

The LDOS in the entire slab and for the full band width is presented in  Fig.~\ref{fig:ldos}(a), while Fig.~\ref{fig:ldos}(b) focuses on subgap energies. In Fig.~\ref{fig:ldos_cuts}(a), cuts of the LDOS between sites 1 and 50 are shown. The LDOS displays the zero-energy Andreev surface bound states, corresponding to the flat band in Fig.~\ref{fig:eigenvalues}. The Andreev bound states lie on sites with odd numbers $a$ \footnote{In graphene zigzag nanoribbons, the zero-energy edge states only lie on one sublattice, i.e. every other site of the full lattice. \cite{FujWakNak96}}. The spectral weight is suppressed near the surface by the Gutzwiller factor $g_a^\mathcal{N}$, see Fig.~\ref{fig:ldos_cuts}(b), as the edge is charged. The suppression of the Gutzwiller factor $g_a^\mathcal{N}$ reflects the strengthened correlations near the edge at the considered low hole doping $\delta=0.05$.


\subsection{Approximation for homogeneous electron densities}\label{sec:ResultsHomogeneous}
\begin{figure}[t!]
    \centering
    \includegraphics[width=\linewidth]{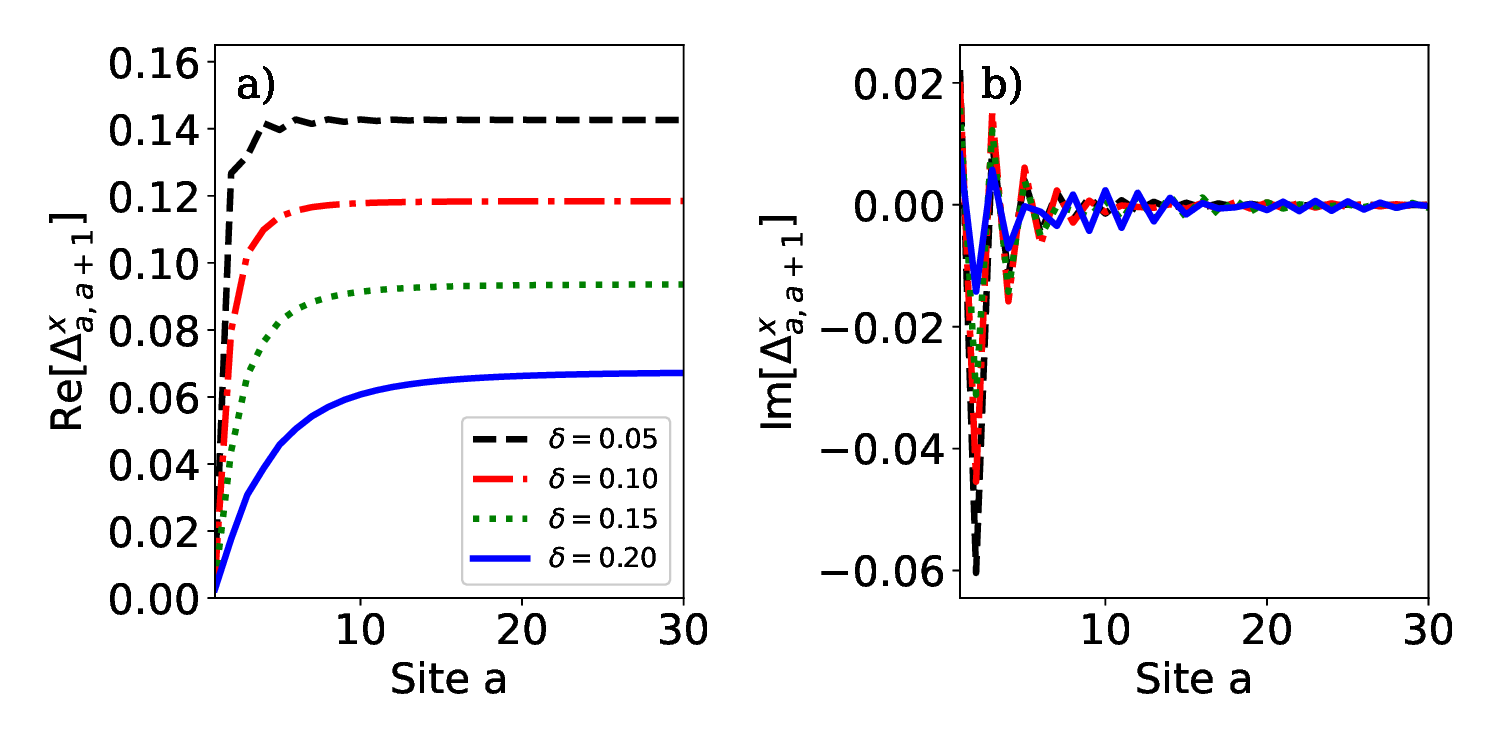}
    \caption{The real (a) and imaginary (b) parts of the pairing amplitude $\Delta_{aa+1}^{x}$ for varying slab hole doping under the assumption that the bulk values for electron density and hopping mean field are the same throughout the slab. The real and imaginary parts have $d$-wave and extended $s$-wave symmetries, respectively. These results agree well with Ref.~\cite{TanTanOga99}.}
    \label{fig:FieldTanuma}
\end{figure}
In this section, we report results for a simplified model, where only pairing amplitude is spatially dependent, while the electron density $n=1-\delta$ and hopping mean field $\chi$ are assumed to be constant throughout the slab. The Lagrange multipliers reflect this. The hopping mean field is taken to have its corresponding bulk value reported in Fig.~\ref{fig:MomentumSpace}, and the Gutzwiller factors are spatially independent and follow from the bulk electron density. Results for this approximative approach was previously reported by Tanuma et~al.~\cite{TanTanOga99}, although our approach is slightly different in that we use the SGA of the t-J model. 
The pairing amplitudes obtained in the SGA are plotted in Fig.~\ref{fig:FieldTanuma}. In contrast to the results of the model allowing for redistribution of electron density presented above, the real part of the pairing amplitude is now suppressed at the edge for all occupations $\delta$, see Fig.~\ref{fig:FieldTanuma}(a). Furthermore, the suppression is more significant, and the pairing amplitude is almost vanishing at the edge, in correspondence with the superconducting $d$-wave order parameter in weak coupling theory. The suppression of the real part, which follows $d$-wave symmetry, Re[$\Delta^x$]=$-$Re[$\Delta^y$], allows for an imaginary part at the edge with extended $s$-wave symmetry Im[$\Delta^x$]=Im[$\Delta^y$] to form, see Fig.~\ref{fig:FieldTanuma}(b).

        

However, as we have seen in previous sections, redistribution of charge and renormalization towards more strong correlations near the edge, leads to a more robust $d$-wave pairing amplitude $\Delta_{a,a+1}^x$ and we find that the extended $s$-wave component is no longer favorable to form.

\section{Summary}\label{sec:Summary}
Slab systems with edges cut at $45^o$ relative to the main crystallographic axes have been studied within the statistically consistent Gutzwiller approach. For uniform electron distribution, the pair-breaking edge reduces the strength of the $d$-wave pairing amplitude and thus allows for an extended $s$-wave component to arise near the edge, as found before \cite{TanTanOga99}. Allowing charge redistribution, charge is attracted by the edge and local electron densities $n_a$ for low average hole-dopings $\delta$ approach one at the edge. As the electron density increases, the Gutzwiller factor $g^t$ is decreased near the edge: a sign of strengthened correlations. At hole doping $\delta=0.2$, the $d$-wave pairing amplitude is reduced near the edge, but less so than in the case of uniform charge distribution. In the underdoped region, when $\delta=0.05$, it is even increased near the edge. 
It should be pointed out, however, that the $d$-wave superconducting order parameter in the correlated state, Eq.~\eqref{eq:order_parameter}, is still reduced at the edge for all doping levels $\delta$ due to the reduction of $g^t$. A reduction which also significantly scales down the spectral weight of zero-energy Andreev bound states in the LDOS. These results may be of importance for the intriguing physics and competition between different symmetry broken states at edges of strongly correlated $d$-wave superconductors.

\acknowledgements
It is a pleasure to acknowledge valuable discussions with J. Spa\l{}ek, M. Fidrysiak, P. Kuterba, K. M. Seja, B. Andersen, M. Randeria, N. Trivedi, and D. Chakraborty. We acknowledge financial support from Chalmers' Area of Advance Nano under its Excellence PhD program, and the Swedish research council, Vetenskapsr\r{a}det, under contract No. 2023-05112. The computations and data handling were enabled by resources provided by the National Academic Infrastructure for Supercomputing in Sweden (NAISS) at C3SE and NSC, partially funded by the Swedish Research Council through Grant Agreement No. 2022-06725. We further acknowledge NAISS for awarding this project access to the LUMI supercomputer, owned by the EuroHPC Joint Undertaking and hosted by CSC (Finland) and the LUMI consortium.
\appendix*

\section{Pair-breaking [110] edge}
The structure of the mean-field Hamiltonian depends on how the edges are oriented relative to the main $xy$-axes of the lattice. Here we consider a slab geometry with $[110]$ edges as shown in Fig.~\ref{fig:Lattice110} and described in Section~\ref{sec:slab}.
The operator $\cd_{i\sigma}$ and $\cnd_{j\sigma}$ appearing in the Hamiltonian Eq.~\eqref{eq:GrandHamiltonian} should be Fourier transformed taking into account translational invariance in the $y'$-direction, keeping the spatial coordinate $x'$ explicilty. To this end, a site $i$ is associated with a pair of real space coordinates $a$ and $b$ specifying its position along the $x'$- and $y'$-axes, respectively. The Fourier transform then takes the form
\begin{align}
    \begin{split}
        \cnd_{ab\sigma} &= \frac{1}{\sqrt{N_y}}\sum_k \cnd_{ak\sigma}e^{ikb}, \\
        \cd_{ab\sigma} &= \frac{1}{\sqrt{N_y}}\sum_k \cd_{ak\sigma}e^{-ikb}.
    \end{split}
\end{align}
The sum over nearest neighbors in this coordinate system has to be taken with care. For example, the position of site 2 can be expressed as $\bm{p}_2=\frac{d}{\sqrt{2}}\hat{x}'$, corresponding to $a=1$, $b=0$. Site 3 on the other hand appears at two different positions in the unit cell, $\bm{p}_3=\frac{d}{\sqrt{2}}(2\hat{x}'\pm \hat{y'})$, corresponding to $a=2$, $b=\pm 1$. The distance between site 2 and 3 in $y'$ direction is accordingly $\pm \frac{d}{\sqrt{2}}$ depending on going in the $x$- or $y$-direction. Using this notation, the sums over nearest neighbors can be expressed as sums of coordinate index pairs $\sum_{\braket{ij}}=\sum_{\braket{aba'b'}}$, i.e. $i\rightarrow ab$, $j\rightarrow a'b'$. The superconducting pairing term acquires the form

\begin{widetext}

\begin{align}
    \begin{split}
\sum_{\braket{ij}\sigma}\lambda^\Delta_{ij}\cnd_{i\sigma}\cnd_{j\sigmab}=
\frac{1}{N_y}    \sum_{\braket{aba'b'}\sigma}\lambda^\Delta_{aba'b'}\cnd_{ab\sigma}\cnd_{a'b'\sigmab}=
\frac{1}{N_y}    \sum_{\braket{aba'b'}\sigma} 
\lambda^\Delta_{aba'b'}\sum_{k k'} \cnd_{ak\sigma}\cnd_{a'k'\sigmab} e^{i(k+k')b} 
e^{ik'(b'-b)}
    =\\
     {\frac{1}{2N_y}}\sum_{aa'\sigma} \left(
\sum_{kk' } \cnd_{ak\sigma}\cnd_{a'k'\sigmab}[(\lambda^{\Delta x}_{aa'}e^{ i\frac{k'd}{\sqrt{2}}}+ 
\lambda^{\Delta y}_{aa'} e^{ -i\frac{k'd}{\sqrt{2}}}) \delta_{a,a'- 1}+(\lambda^{\Delta x}_{aa'}e^{ -i\frac{k'd}{\sqrt{2}}}+ 
\lambda^{\Delta y}_{aa'} e^{ i\frac{k'd}{\sqrt{2}}}) \delta_{a,a'+ 1}]\underbrace{\sum_b e^{i(k+k')b}}_{\delta_{k+k',0}}  \right) =\\
     {\frac{1}{2}}\sum_{aa'\sigma} \left(
\sum_{k } \cnd_{ak\sigma}\cnd_{a'-k\sigmab}[(\lambda^{\Delta x}_{aa'}e^{ -i\frac{kd}{\sqrt{2}}}+ 
\lambda^{\Delta y}_{aa'} e^{ i\frac{kd}{\sqrt{2}}}) \delta_{a,a'- 1}+(\lambda^{\Delta x}_{aa'}e^{ i\frac{kd}{\sqrt{2}}}+ 
\lambda^{\Delta y}_{aa'} e^{ -i\frac{kd}{\sqrt{2}}}) \delta_{a,a'+ 1}]\right)=\\
     {\frac{1}{2}}\sum_{aa'\sigma} \left(
\sum_{k } \cnd_{ak\sigma}\cnd_{a'-k\sigmab}\left[\eta^-_{aa'}(k)\delta_{a,a'-1} + \eta^+_{aa'}(k) \delta_{a,a'+ 1}\right]\right)=\\
     {\frac{1}{2}} \sum_{k\sigma} \left(
\sum_{a=1 }^{N_x-1} \cnd_{ak\sigma}\cnd_{a+1-k\sigmab}\eta^-_{a,a+1}(k) + \sum_{a=2 }^{N_x} \cnd_{ak\sigma}\cnd_{a-1-k\sigmab}\eta^+_{a,a-1}(k) \right)=\\
\sum_{k\sigma}\sum_{ a=1 }^{N_x-1} \cnd_{ak\sigma}\cnd_{a+1-k\sigmab}\eta^-_{a,a+1}(k).
\end{split}
\end{align}
\end{widetext}
When evaluating the sum over $b'$ in the first step, it has been used that $\lambda^\Delta_{aba'b'}$ is independent of position in the $b$-direction due to the translational symmetry. The exponential have been rewritten as
\begin{align}
    \begin{split}
\lambda^\Delta_{aba'b'}\cnd_{ab\sigma}\cnd_{a'b'\sigmab} &= \sum_{k k'} \cnd_{ak\sigma}
e^{ikb} 
\cnd_{a'k'\sigmab} e^{ik'b'} \\
&= \sum_{k k'} \cnd_{ak\sigma}\cnd_{a'k'\sigmab} e^{i(k+k')b} 
e^{ik'(b'-b)}
    \end{split}
\end{align}
to allow the sum over $b$ to be executed. Further, note that the distance between sites $a$ and $a+1$ comes with different signs in the b-direction ($b'-b$), therefore $x$- and $y$-links comes with different signs in the exponential. Lastly, to avoid double counting of links a factor $1/2$ is introduced.

Fourier transforming Eq.~\eqref{eq:GrandHamiltonian} in the same fashion, the terms containing operators reduce to

\begin{align}
\begin{split}
\sum_{a  k \sigma} \Bigl( &
4\lambda^{\chi}_{aa+1}(\cd_{ak\sigma}\cnd_{a+1k\sigma}+\cd_{a+1k\sigma}\cnd_{ak\sigma})\cos{\frac{kd}{\sqrt{2}}} \\
&+\cnd_{ak\sigma}\cnd_{a+1-k\sigmab}\eta^-_{a,a+1}(k)\\
&+ \cd_{a+1-k\sigmab}\cd_{ak\sigma}\eta^-_{a,a+1}(k)^*\\
&+ (\mu_a+\lambda_{a}^n )\cd_{ak\sigma}\cnd_{ak\sigma} \Bigr),
 \label{eq:operator_term}
\end{split}
\end{align}
which can be written as $\sum_k \hat{\Psi}_k^\dagger \bm{H}_k \hat{\Psi}_k$, where $\bm{H}_k $ and $\hat{\Psi}_k$ are defined in the main text. 

Next the mean-field approximation of the expectation value of the effective Hamiltonian, $W$ in Eq.~\eqref{eq:effective_energy}, is evaluated. 
The paring term takes the form 
\begin{widetext}
\begin{align}
\begin{split}
\sum_{\braket{ij}}  Jg_{ij}^s\frac{3}{2}|\Delta_{ij}|^2
&= {\frac{1}{2}}\sum_{aa'\braket{bb'}\pm}  Jg_{aa'}^s\frac{3}{2} |\Delta_{aba'b'}|^2\delta_{a,a'\pm1}
= \frac{3JN_y}{4}  \sum_{f\in\{x,y\}}  \left[\sum_{a=1}^{N_x-1} g_{aa+1}^s|\Delta_{aa+1}^f|^2+\sum_{a=2}^{N_x}g_{aa-1}^s|\Delta_{aa-1}^f|^2 \right]\\
&=
\frac{3JN_y}{2}  \sum_{a=1}^{N_x-1} g_{aa+1}^s(|\Delta_{aa+1}^x|^2+|\Delta_{aa+1}^y|^2),
\end{split}
\end{align}
\end{widetext}
where, as before, the paring on $x$ and $y$ links are differentiated to allow for both extended $s$- and $d$-wave symmetries, and a $\frac{1}{2}$ is introduced to avoid double-counting. Adding the hopping term, which is independent of direction, and the conjugate term of the pairing, gives $W_{110}$ in Eq.~\eqref{eq:W110}.

Lastly, the terms with products between mean-fields and lambdas are considered. The pairing becomes 

\begin{align*}
    \begin{split}
\sum_{\braket{ij}\sigma}\lambda^\Delta_{ij}\Delta_{ij}
   &=2\sum_{\braket{aba'b'}} \lambda^\Delta_{aba'b'}\Delta_{aba'b'}\\
   &=2N_y\sum_{\braket{aa'}} \left( \lambda^{\Delta x}_{aa'}\Delta_{aa'}^x+\lambda^{\Delta y}_{aa'}\Delta_{aa'}^y
   \right)\\
   &=2N_y\sum_{a=1}^{N_x-1} \left( \lambda^{\Delta x}_{aa+1}\Delta_{aa+1}^x+\lambda^{\Delta y}_{aa+1}\Delta_{aa+1}^y
   \right),
\end{split}
\end{align*}
and the full contribution is
\begin{align}
    \begin{split}
{\lambda_{110}} &= 2N_y\sum_{a=1}^{N_x-1} \Bigl( 4\lambda^{\chi }_{a,a+1}\chi_{a,a+1}\\
&\hspace{1.9cm}+\lambda^{\Delta x}_{a,a+1}\Delta_{a,a+1}^x + \lambda^{\Delta y}_{a,a+1}\Delta_{a,a+1}^y\\
&\hspace{1.9cm}+\lambda^{\Delta x*}_{a,a+1}\Delta_{a,a+1}^{x*}+\lambda^{\Delta y*}_{a,a+1}\Delta_{a,a+1}^{y*}\Bigr)\\
&\quad+N_y\sum_a\left( \lambda^n_a n_a -\lambda^n_a-\mu \right).
\label{eq:lambda110}
\end{split}
\end{align}

It is now possible to express the Grand potential functional per site as
\begin{equation}
        \frac{\Omega}{N}=\frac{W_{110}}{N}-\frac{1}{N\beta} \sum_{lk}\ln{(1+e^{-\beta E_{lk}})}+\frac{\lambda_{110}}{N},
\end{equation}
where $N=N_yN_x$ is the number of sites in the lattice, and $E_{lk}$ is the l:th eigenvalue of the Hamiltonian $\bm{H}_k $ for momentum $k$. With the aim of finding the minimum of $\Omega$, derivatives are taken with respect to the mean-fields and $\lambda$ parameters. The derivatives with respect to lambda parameters lead to self-consistency equations

\begin{widetext}

\begin{equation}
\left\{
\begin{aligned}[c]
\pdv{\Omega}{\lambda^\chi_{a,a+1}} = 0\\
\pdv{\Omega}{\Re\lambda^{\Delta^x}_{a,a+1}} = 0\\
\pdv{\Omega}{\Im\lambda^{\Delta^x}_{a,a+1}} = 0\\
\pdv{\Omega}{\lambda^n_{a}} = 0
\end{aligned}\right.
\qquad\Rightarrow\qquad
\left\{
\begin{aligned}[c]
\chi_{a,a+1} &= -\frac{1}{8N_y}\sum_{lk}f(E_{lk})\pdv{E_{lk}}{\lambda^\chi_{a,a+1}}\\
\Re\Delta_{a,a+1}^x &= -\frac{1}{4N_y}\sum_{lk}f(E_{lk})\pdv{E_{lk}}{\Re\lambda^{\Delta^x}_{a,a+1}} \\
\Im\Delta_{a,a+1}^{x} &= \frac{1}{4N_y}\sum_{lk}f(E_{lk})\pdv{E_{lk}}{\Im\lambda^{\Delta^x}_{a,a+1}}\\
n_{a} &= 1 - \frac{1}{N_y}\sum_{lk}f(E_{lk})\pdv{E_{lk}}{\lambda^n_{a}}
\end{aligned}\right.
\label{eq:mean_fields}
\end{equation}
where $f(E)$ is the Fermi-Dirac distribution function. Derivatives of $\Omega$ with respect to mean fields give corresponding lambda parameters 
\begin{equation}
\left\{
\begin{aligned}[c]
\pdv{\Omega}{\chi_{a,a+1}} = 0\\
\pdv{\Omega}{\Re\Delta_{a,a+1}^{x}} = 0\\
\pdv{\Omega}{\Im\Delta_{a,a+1}^{x}} = 0
\end{aligned}\right.
\qquad\Rightarrow\qquad
\left\{
\begin{aligned}[c]
\lambda_{a,a+1}^\chi &= -tg_{a,a+1}^t + \frac{3}{4}Jg^s_{a,a+1}\chi_{a,a+1}\\
\Re\lambda_{a,a+1}^{\Delta^x} &= \frac{3}{4}Jg_{a,a+1}^s\Re\Delta_{a,a+1}^{x} \\
\Im\lambda_{aa+1}^{\Delta^x} &= -\frac{3}{4}Jg_{a,a+1}^s\Im\Delta_{a,a+1}^{x}
\end{aligned}\right.
\label{eq:lambdas}
\end{equation}
while the derivative with respect to electron density gives
\begin{align}
\begin{split}
\pdv{\Omega}{n_{a}} = 0
\qquad\Rightarrow\qquad
\lambda^n_{a} &=
-8t\pdv{g_{a,a+1}}{n_a}\chi_{a,a+1} 
+\frac{3}{2}J\pdv{g_{a,a+1}^s}{n_a}\left(2|\chi_{a,a+1}|^2+|\Delta_{a,a+1}^x|^2 + |\Delta_{a,a+1}^y|^2\right)\\
&\quad -8t\pdv{g_{a-1,a}}{n_a}\chi_{a-1,a}
+\frac{3}{2}J\pdv{g_{a-1,a}^s}{n_a}\left(2|\chi_{a-1,a}|^2+|\Delta_{a-1,a}^x|^2+|\Delta_{a-1,a}^y|^2\right).
\end{split}
\label{eq:lambda_n}
\end{align}
%
%
\end{widetext}
Since $\Delta^x=\Re\Delta^{x}+i\Im\Delta^{x}$ and $\lambda^\Delta=\Re\lambda^{\Delta^x}+i\Im\lambda^{\Delta^x}$ may be complex, derivatives are taken with respect to the real and imaginary parts separately. The equations for $y$-links are obtained by changing superscript $x$ to $y$. 

The above equations are solved until self-consistency has been achieved to a maximum relative accuracy of $10^{-4}$ of any quantity at any lattice site. Self consistent solutions of mean fields and Lagrange multipliers are obtained using 128 $k$-values (equal to number of unit cells in the $y$-direction) while the local density of states are found by using 512 $k$-values.  Our code is parallelized using the \textit{MPI.jl} package \cite{MPI}.

\bibliography{bibliography}

\end{document}